\documentclass{optica-article}

\journal{optcon}

\articletype{Research Article}

\usepackage{lineno}

\begin{document}

\title{Topological spin defects of light}

\author{Haiwen Wang,\authormark{1,+} Charles C. Wojcik,\authormark{2} and Shanhui Fan\authormark{2,*}}

\address{\authormark{1}Department of Applied Physics, Stanford University, Stanford, CA, 94305, USA\\
\authormark{2}Department of Electrical Engineering, Stanford University, Stanford, CA, 94305, USA}

\email{\authormark{+}hwwang@stanford.edu}
\email{\authormark{*}shanhui@stanford.edu} 



\begin{abstract}
Topological defects are found in a variety of systems, and their existence are robust under perturbations due to their topological nature. Here we introduce a new type of topological defects found in electromagnetic waves: topological spin defects. Such a defect is associated with a point where the electromagnetic spin density is zero, and generically have a nontrivial topological spin texture surrounding the defect point. Due to such spin texture, a topological spin defect possesses a quantized topological charge. We provide examples where isolated defect point, periodic or quasi-periodic defect lattices can be created. Such topological spin defect points may find applications in 3D imaging and nanoparticle manipulation.
\end{abstract}

\section{Introduction}
Topological defects are topological singularities in the distributions of the physical fields in a system. These singularities carry topological charges and hence their existence is robust under small perturbations \cite{mermin1979topological}. Initially developed in condensed matter physics \cite{harada1992real, zwierlein2005vortices, ray2014observation, nikkhou2015light, xu2015discovery, yazyev2010topological, milde2013unwinding, fujishiro2019topological, da2014observation}, in recent years there has been significant interests in exploring topological defects in photonics. Many important photonic effects, including bound state in the continuum \cite{zhen2014topological, yin2020observation}, photonic Weyl point \cite{xiao2016hyperbolic}, phase and polarization singularities \cite{leach2005vortex, song2018broadband, wang2021engineering, flossmann2005polarization, zhan2009cylindrical, dennis2009singular}, and exceptional point \cite{wang2019dynamics, ozdemir2019parity}, can all be described in terms of a topological defect of certain optical fields in real, momentum, or parameter space. These singularities have found many applications in imaging \cite{dorn2003sharper, yan2015q}, optical communication \cite{bozinovic2013terabit}, and sensing \cite{lai2019observation}. 

In this paper we report a new type of topological defect in photonics, the topological spin defect. We show that the distribution of the spin angular momentum (SAM) density of an electromagnetic field in three dimensions (3D) may contain topological singularities, which we refer to as a topological spin defect. The spin angular momentum density of a monochromatic electromagnetic field at a specific spatial location is defined as \cite{du2019deep, dai2020plasmonic, lei2021photonic, zhang2022optical, dai2022ultrafast, lin2021photonic}:
\begin{equation}
    \mathbf{S} = \frac{1}{4\omega}\Bigl[\epsilon_0\mathrm{Im}(\mathbf{E^*}\times \mathbf{E}) + \mu_0\mathrm{Im}(\mathbf{H^*}\times \mathbf{H})\Bigr]
    \label{Sdef}
\end{equation}
where $\omega$ is the angular frequency, $\epsilon_0$ and $\mu_0$ are the vacuum permittivity and permeability, $\mathbf{E}$ and $\mathbf{H}$ are the electric and magnetic field vectors, respectively. A topological spin defect corresponds to a point where $\mathbf{S} = 0$. We show that generically, for every sphere in the vicinity of and enclosing such a point, the spin distribution on the sphere exhibits a skyrmion texture. An example of such texture is shown in Fig.~\ref{Fig1}d. The simplest skyrmion texture is a monopole texture. More generally, a skyrmion texture can be related to a monopole or an anti-monopole texture by applying spatially dependent but continuous rotations. Such topological spin defects were previosusly discovered in chiral magnets, with various names such as hedgehogs, Bloch points, or emergent monopoles \cite{milde2013unwinding, da2014observation, fujishiro2019topological, kanazawa2020direct, gobel2021beyond}. We, for the first time, points to their existence in the spin density of electromagnetic fields. Compare with condensed matter systems, topological spin defects of light is directly controlled by the electromagnetic field distribution, and can be easily reconfigured into various locations, charges or spin textures.


Our work is related to but differ from all previous works on the study of topological properties of electromagnetic field distributions. In particular, skyrmion texture has been previously explored as two dimensional distributions of photon spins in a number of systems \cite{du2019deep, dai2020plasmonic, lei2021photonic, zhang2022optical, dai2022ultrafast, lin2021photonic, van2019photonic}. In two dimensions, one can prove that the spin texture in general cannot exhibit a topological defect \cite{mermin1979topological}. And indeed, in all these works on skyrmion texture in two dimensions, the spin does not exhibit an isolated zero point. More generally, interesting topological features have been discovered in the polarization \cite{wang2021polarization, gao2020paraxial, guo2020meron, shen2021topological}, field vector \cite{shen2021supertoroidal, deng2022observation, tsesses2018optical, davis2020ultrafast}, or phase properties \cite{allen1992orbital, dennis2009singular} of electromagnetic fields. Our work differ from these works in that we focus on the spin properties.  Exploration of the topological properties of photon spins is of interest since the spin properties of photons can directly influence light-matter interactions \cite{dietrich1994high, canaguier2013force, bliokh2014magnetoelectric, goldman2014light, zhang2005induced}.

\section{General properties of spin defect}
We start by first describing the general properties of the spin defect point in electromagnetic fields. The spin defect corresponds to a point $\mathbf{R}_0$ where the spin density given by Eq. (\ref{Sdef}) is zero. We assume that the electric and magnetic fields of the monochromatic light field at angular freuquency $\omega$ is $\mathbf{E}(\mathbf{R})$ and $\mathbf{H}(\mathbf{R})$ respectively, and we expand the fields locally around $\mathbf{R}_0$:
\begin{equation}
\begin{split}
    \mathbf{E}(\mathbf{r}) &= \mathbf{E_0} + \bar{\mathbf{e}}\cdot \mathbf{r} + O(r^2) \\
    \mathbf{H}(\mathbf{r}) &= \mathbf{H_0} + \bar{\mathbf{h}}\cdot \mathbf{r} + O(r^2)
    \label{field_expansion}
\end{split}
\end{equation}
where $\mathbf{E_0}$ and $\mathbf{H_0}$ are respectively the electric and magnetic fields at $\mathbf{R_0}$, $\mathbf{r}=\mathbf{R} - \mathbf{R}_0$, $\bar{\mathbf{e}}\equiv \mathbf{\nabla E}$ and $\bar{\mathbf{h}}\equiv \mathbf{\nabla H}$ are both second order tensors. Inserting Eq. (\ref{field_expansion}) back into Eq. (\ref{Sdef}), and noting that the spin at $\mathbf{R_0}$ is zero, we get:
\begin{equation}
    \mathbf{S} = \bar{\mathbf{A}}\cdot \mathbf{r} + O(r^2)
    \label{localS}
\end{equation}
The second order tensor
\begin{equation}
    \bar{\mathbf{A}}=\frac{1}{2\omega}\Bigl[\epsilon_0\mathrm{Im}(\mathbf{E_0^*\times\bar{\mathbf{e}}})+\mu_0\mathrm{Im}(\mathbf{H_0^*\times\bar{\mathbf{h}}})\Bigr]
\end{equation}
It can also be viewed as a $3\times 3$ real matrix. When $\bar{\mathbf{A}}$ has full rank, which is the generic case, we get an isolates zero of spin density at $\mathbf{r}=0$.

The topological property of the spin distribution is manifested in the topological charge \cite{han2017skyrmions, milde2013unwinding, fujishiro2019topological}:
\begin{equation}
    Q = \frac{1}{8\pi}\int_\Sigma \epsilon_{ijk} \mathbf{n}(\mathbf{r})\cdot [\partial_i\mathbf{n}(\mathbf(r))\times\partial_j\mathbf{n}(\mathbf{r})] \mathrm{d}\sigma_k
    \label{Qdef}
\end{equation}
where $\mathbf{n} = \mathbf{S}/|\mathbf{S}|$ is the normalized spin vector, $\epsilon$ is the Levi-Civita tensor, $\mathrm{d}\boldsymbol\sigma$ is a vector of infinitesimal surface element along the surface normal, the integration area $\Sigma$ is a sphere surrounding the spin defect point. The integrand is known as the skyrmion density, topological current, or emergent magnetic field \cite{han2017skyrmions, milde2013unwinding, fujishiro2019topological}. (There is in fact a connection of such topological charge to the Chern number, as discussed in the supplementary section 2.) For $\mathbf{S}$ as described by Eq. (\ref{localS}), the integration in Eq. (\ref{Qdef}) results in
\begin{equation}
    Q=\mathrm{sign}(\mathrm{det}\bar{\mathbf{A}})
\end{equation}
and hence $Q=\pm 1$ when $\bar{\mathbf{A}}$ is non-singular. As an illustration, when $\bar{\mathbf{A}} = \pm I_{3\times 3}$, the spin distribution given by Eq. (\ref{localS}) corresponds to a standard monopole with charge $\pm 1$. General non-singular matrix $\bar{\mathbf{A}}$ can be continuously transformed into either $I_{3\times 3}$ or $-I_{3\times3}$ while maintaining fully ranked during transformation. Therefore a spin distribution with a non-singular $\bar{\mathbf{A}}$ has either $+1$ or $-1$ topological charges. We provide a construction of such transformations in the supplementary section 1. As an alternative derivation for the topological properties of such spin distribution, the mapping from a 2-sphere (here being the spherical surface $\Sigma$ in real space) to a 2-sphere (here being the space of normalized spin vector $\mathbf{n}$) is classified by the second homotopy group $\pi_2(S^2)$ of the 2-sphere. This group being nontrivial allows the existence of topological charges.

\section{Spin defect point in superposed Gaussian beams}
\begin{figure}
    \includegraphics[width=0.75\columnwidth]{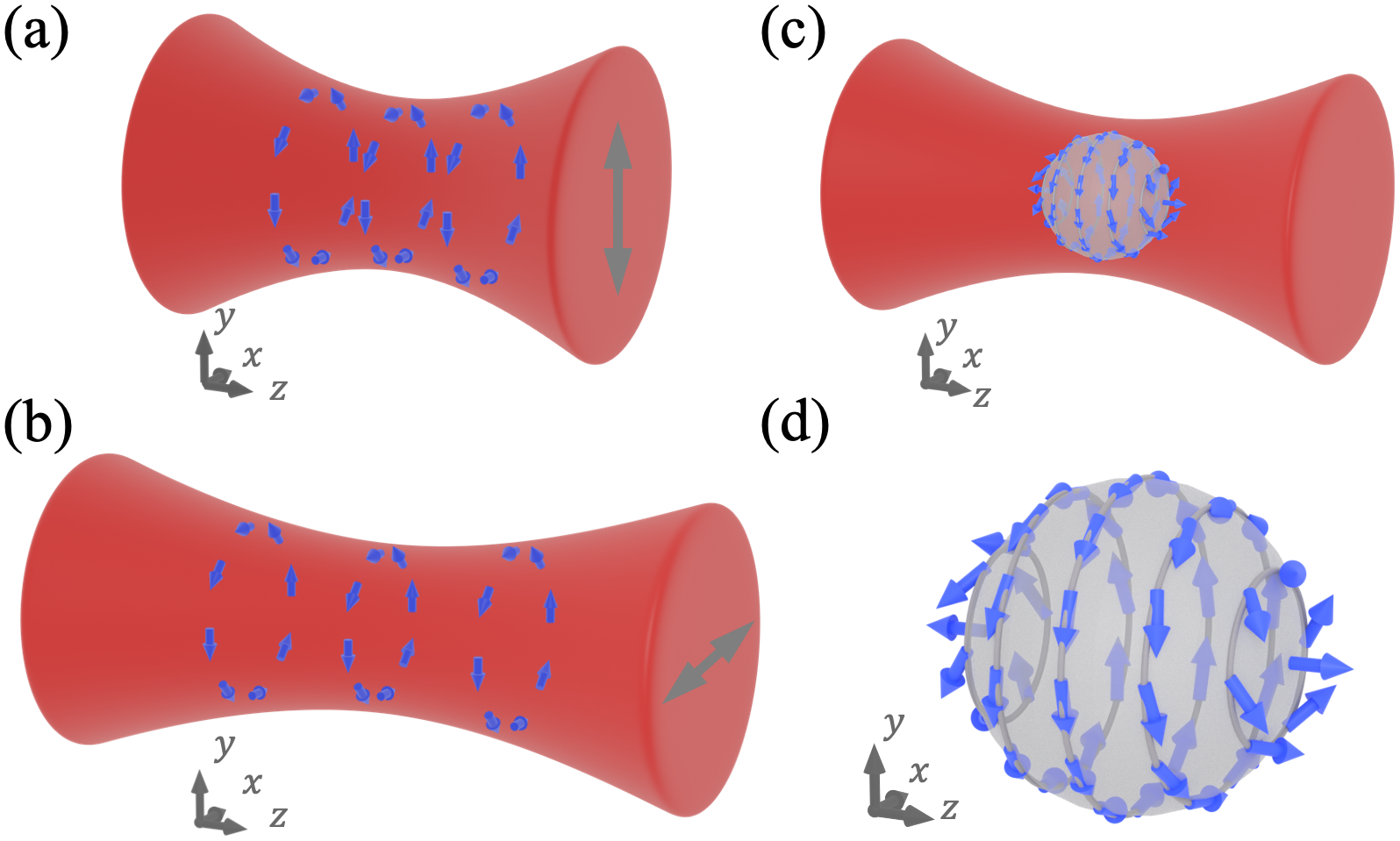}%
    \centering
    \caption{Various spin configurations in Gaussian beams and superposition of Gaussian beams. (a) The normalized spin density vectors for a $y$-polarized Gaussian beam with $z_R=45\lambda$. The spin are calculated on three planes with $z=-2.5\lambda, 0, 2.5\lambda$. (b) The normalized spin density vectors for an $x$-polarized Gaussian beam with $z_R=90\lambda$. The spin are calculated on three planes with $z=-4\lambda, 0, 4\lambda$. (c) Superposition of the beam in (a) and (b) give rise to a topological spin defect at the focal point. The normalized spin vectors are evaluated on a sphere with radius $r=2\lambda$ at the focal point. (d) Magnified view of the spin vectors shown in (c). The Gaussian profile in (a-c) are for illustration purpose and not plotted in actual scale. \label{Fig1}}
\end{figure}

We now proceed to demonstrate how topological spin defects can be created in actual electromagnetic fields. We first examine the well known Gaussian beam \cite{siegman1986lasers, april2010ultrashort}. Previous researches have shown that linearly polarized Gaussian beam, along with other tightly focused beams or guided waves that are linearly polarized, give rise to a photon spin distribution that is largely transverse to the beam propagation direction \cite{aiello2015transverse, neugebauer2015measuring, bekshaev2015transverse, shi2021spin}. We reproduce these results in Fig.~\ref{Fig1}a and b. In our modeling, we start by a scalar field distribution with a Gaussian beam profile:
\begin{linenomath}
\begin{equation}
    u(x,y,z)=\frac{N}{z-iz_R}\exp{\left[\frac{ik(x^2+y^2)}{2(z-iz_R)}+ikz\right]}
    \label{GB}
\end{equation}
\end{linenomath}
where $k=2\pi/\lambda$, $\lambda$ is the wavelength of light, $z_R$ is the Rayleigh range, $N$ is a normalization constant. To obtain the electric and magnetic field components, we use the Hertz potential method \cite{jackson1999classical, april2010ultrashort}, where the fields are related to the Hertz potential $\mathbf{\Pi_m}$ by:
\begin{equation}
    \begin{split}
        \mathbf{E} &= i\omega\mu_0\boldsymbol\nabla\times\mathbf{\Pi_m} \\
        \mathbf{H} &= \boldsymbol\nabla\times\boldsymbol\nabla\times\mathbf{\Pi_m}
        \label{HP}
    \end{split}
\end{equation}
To obtain a $y$-polarized Gaussian beam (which has no $x$ component electric field everywhere), we assume that the $x$ component of the Hertz potential $\mathbf{\Pi_m}$ is given by the Gaussian beam profile in Eq. (\ref{GB}), and all other components are zero. The spin distribution as calculated using Eq. (\ref{Sdef}) is then plotted in Fig.~\ref{Fig1}a. We observe that at the beam waist, the spin is exactly in the transverse direction with no $z$ component. In the positions around the beam waist, the spin component contain small $z$ components but are too small to be seen visually in the figure. The spin is exactly zero along the axis of the beam, since the fields are linearly polarized on axis for both electric and magnetic fields. (We note that the axis here is not a topological spin defect. The topological spin defect corresponds to the vanishing of spin at an isolated point in space.) Similarly, we can obtain an $x$-polarized beam by choosing the Hertz potential $\mathbf{\Pi_m}$ only having $y$ component being the Gaussian beam profile. Its spin density is illustrated in Fig.~\ref{Fig1}b.

We note that the treatment here is based on paraxial approximation. We have numerically verified the same results hold in the exact solution of the Maxwell's equations. In particular, our description based on the Hertz potential provides an accurate description of the longitudinal fields that are crucial for the description of the transverse spin. We provide more details in the supplementary section 3.

As the first main result of our work, we now show that an appropriate superposition of a $y$-polarized Gaussian beam (Fig.~\ref{Fig1}a) with an $x$-polarized Gaussian beam (Fig.~\ref{Fig1}b) creates a topological spin defect in 3D real space. The Hertz potential for the superposed beam is:
\begin{equation}
    \mathbf{\Pi_m}(x,y,z) = 
    \begin{pmatrix}
    \frac{N_1}{z-iz_{R1}}\exp{\left[\frac{ik(x^2+y^2)}{2(z-iz_{R1})}+ikz\right]} \\
    \frac{N_2}{z-iz_{R2}}\exp{\left[\frac{ik(x^2+y^2)}{2(z-iz_{R2})}+ikz\right]} \\
    0
    \end{pmatrix}
    \label{superposition}
\end{equation}
Here $N_1$ and $N_2$ are normalization constants, $z_{R1}$, $z_{R2}$ are two Rayleigh ranges. The two beams both propagate in $z$ direction and share a common focus at $x=y=z=0$. When $z_{R1}\neq z_{R2}$, and $N_1$, $N_2$ being nonzero real constants, the Hertz potential in Eq. (\ref{superposition}) give rise to a topological spin defect at focus. As illustrated in Fig.~\ref{Fig1}c and d, we choose $z_{R1}=45\lambda$, $z_{R2}=90\lambda$, and $N_1=N_2$.

At the focal point, the field is linearly polarized and the spin density is zero. We show that the focal point is a topological spin defect. In Fig.~\ref{Fig1}a and b, we see that on the waist and away from the focal axis, the spin loops around the axis. This feature remains the same for the superposed beam as shown in Fig.~\ref{Fig1}c. On the other hand, on axis but away from the focus, the two Gaussian beam has different Rayleigh ranges and therefore different Gouy phases. The Gouy phase difference makes the polarization elliptical on axis, for both electric and magnetic fields. Such phase differences are opposite on opposite sides of the focal point. The $z$ component of the spin therefore has opposite signs on opposite sides of the focus. With these effects considered together, we get a spin defect point at the focal point and a topologically nontrivial spin configuration around such point. This is in agreement with our general mathematical discussion above which shows that an isolated zero of the spin density is generically topological.

\begin{figure}
    \includegraphics[width=0.7\columnwidth]{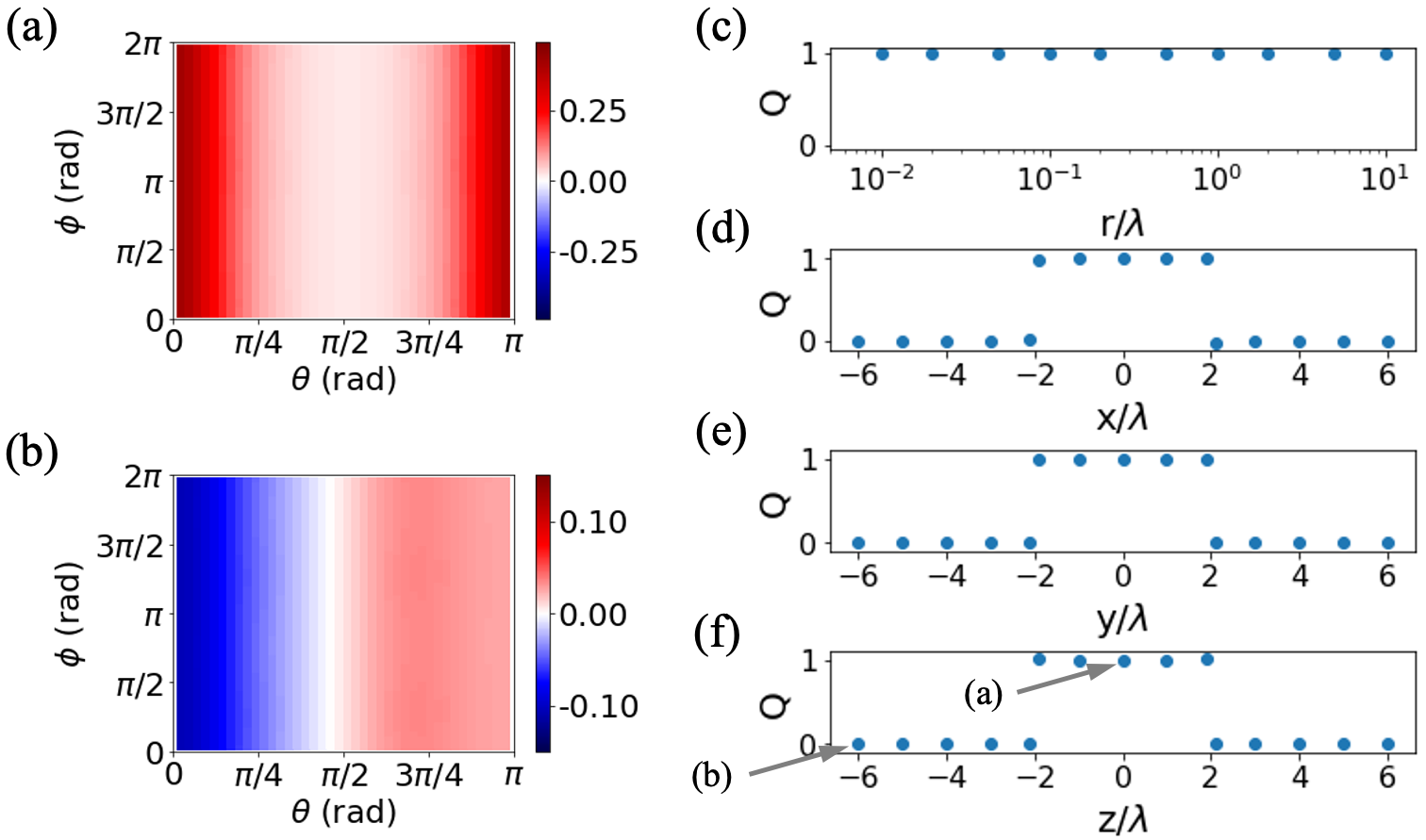}%
    \centering
    \caption{Calculation of the topological charge of for the two superposed beams as shown in Fig.~\ref{Fig1}c and d, which exhibits a topological spin defect centered at $x = y = z = 0$.  (a) Skyrmion density $\mathcal{F}$ on the sphere with radius $r=2\lambda$ surrounding the focal point at $x=y=z=0$. (b) Skyrmion density $\mathcal{F}$ on a sphere with same radius but centered at $x=y=0, z=-6\lambda$. (c) Integrated topological charge for spheres with different radius $r$ centered at focal point. (d-f) Integrated topological charge for spheres with radius $r=2\lambda$ and displaced in (d) $x$, (e) $y$, and (f) $z$ directions. \label{Fig2}}
\end{figure}

We calculate the topological charge of the spin defect shown in Fig.~\ref{Fig1}c and d by applying Eq. (\ref{Qdef}) on a sphere of radius $r=2\lambda$. In spherical coordinates, the skyrmion density is $\mathcal{F}=\frac{1}{4\pi\sin\theta}\mathbf{n}\cdot\left(\frac{\partial\mathbf{n}}{\partial\theta}\times\frac{\partial\mathbf{n}}{\partial\phi}\right)$, where $\theta$ is the polar angle measured from the $+z$ axis, $\phi$ is the azimuthal angle measured from the $+x$ axis. The value of the skyrmion density is non-negative everywhere as plotted in Fig.~\ref{Fig2}a, and its integration gives $Q=1$. On the other hand, the same sphere centered on axis with $z=-6\lambda$ has the skyrmion density shown in Fig.~\ref{Fig2}b. The skyrmion density contains both positive and negative values and integrates to zero. Thus, the total charge enclosed in such sphere is zero. Furthermore, the spin component $S_z$ is always negative when $z<0$, therefore no topological spin defect exists in a sphere residing entirely in the $z < 0$ half space.

In Fig.~\ref{Fig2}c, we apply Eq. (\ref{Qdef}) on a series of spheres centered at the focal point with different radii $r$. The integrated topological charge is all close to one, due to the contribution of the spin defect at the focus. In Fig.~\ref{Fig2}d, e, and f, we choose the sphere radius to be $r=2\lambda$ and displace it in $x$, $y$, and $z$ directions away from the focus. We see the topological number changes abruptly to zero when the displacements are larger than $2\lambda$, where the spin defect at the origin moves out of the sphere. These numerical results shows that this spin defect is an isolated singularity, carries $+1$ topological charge, and there is no other spin defect points nearby.

\section{Spin defect lattices in interfering plane wave systems}
\begin{figure}
    \includegraphics[width=0.7\columnwidth]{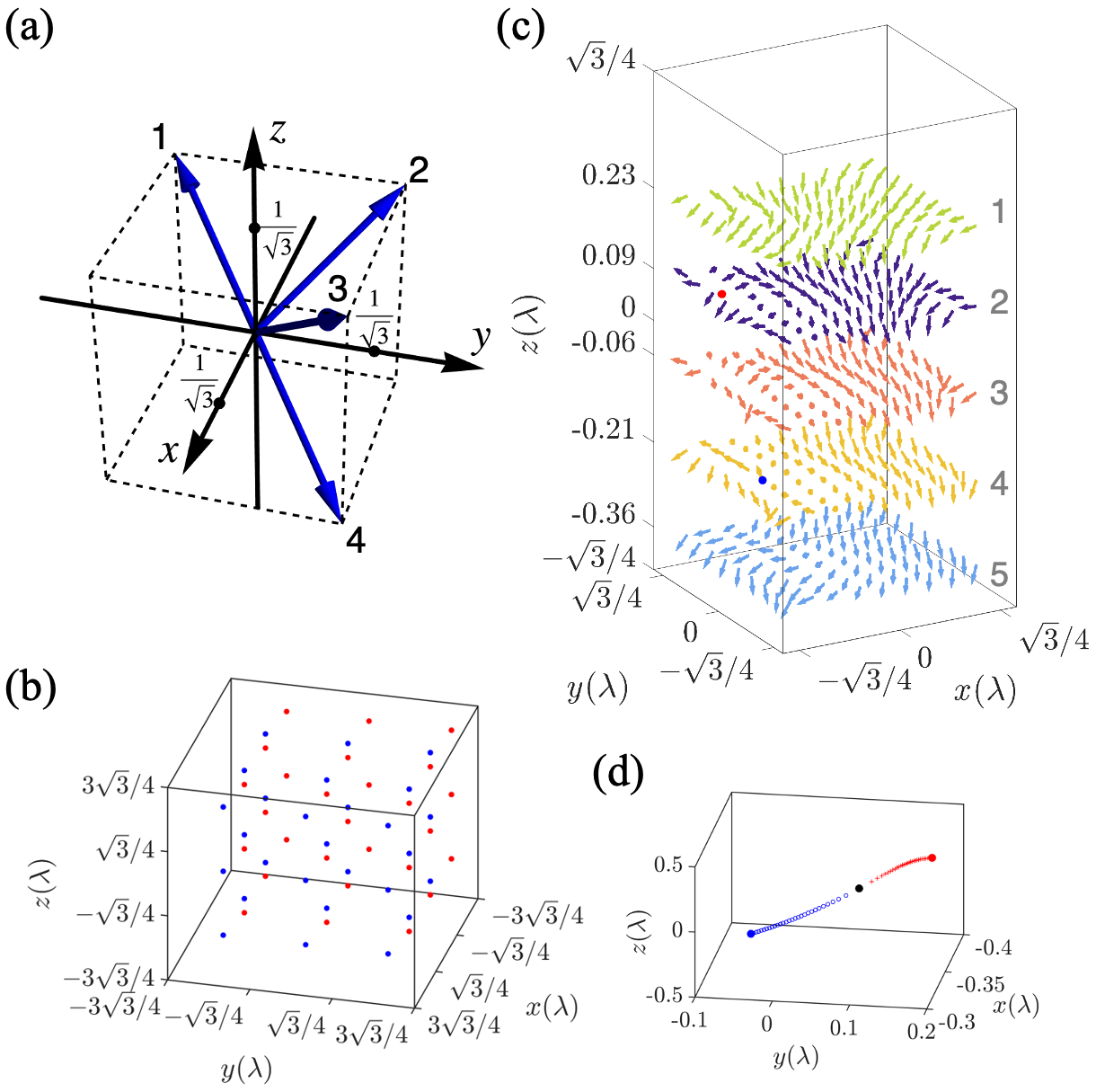}%
    \centering
    \caption{Spin defect lattices created by interfering plane waves. (a) The propagation direction $\hat{\mathbf{k}_i}$ for four plane waves. (b) The location of positive (red dot) and negative (blue dot) topological charges for the spin distribution as generated by the four plane waves shown in (a). (c) Distribution of the normalized spin vector inside one unit cell, evaluated on five planes with different values in $z$, numbered 1-5. Plane 2 and 4 coincide with the $z$-positions of positive charge (red dot) and negative charge (blue dot) respectively. (d) Location of the positive and negative charge inside one unit cell when the amplitude of plane wave 3 is increased. Black dot mark the point when they annihilate. \label{Fig3}}
\end{figure}

In addition to a single topological spin defect as discussed in the last section, many complex configurations of topological spin defects can be found in electomagnetic waves. In the following, we provide an example that interfering planewaves can give rise to 3D spin defect lattices. It was known that in 3D, the interference of four non-coplanar plane waves give rise to a 3D periodic intensity pattern \cite{lu2010interference}. Therefore, to construct a 3D lattice of topological spin defects, we consider the superposition of four plane waves with the resulting electric and maganetic field described by:
\begin{equation}
\begin{split}
    \mathbf{E}(\mathbf{R}) &= \sum_{i=1,2,3,4} (A_{s,i}\mathbf{e}_{s,i}+A_{p,i}\mathbf{e}_{p,i})\mathrm{e}^{i\mathbf{k}_i\cdot\mathbf{R}} \\
    \mathbf{H}(\mathbf{R}) &= \frac{1}{Z_0}\sum_{i=1,2,3,4} (-A_{s,i}\mathbf{e}_{p,i}+A_{p,i}\mathbf{e}_{s,i})\mathrm{e}^{i\mathbf{k}_i\cdot\mathbf{R}}
    \label{fourplanewave}
\end{split}
\end{equation}
Here $A_{s,i}$ ($A_{p,i}$) and $\mathbf{e}_{s,i}$ ($\mathbf{e}_{p,i}$) are the $s$ ($p$) polarization amplitude and unit polarization vector of the $i$-th plane wave, respectively. $\mathbf{k}_i$ is the wavevector of the $i$-th plane wave. $Z_0$ is vacuum impedance. We choose the directions of the wavevectors $\hat{\mathbf{k}_i}$ as $\frac{1}{\sqrt{3}}(-1,-1,1)^\mathrm{T}$, $\frac{1}{\sqrt{3}}(-1,1,1)^\mathrm{T}$, $\frac{1}{\sqrt{3}}(1,1,1)^\mathrm{T}$, $\frac{1}{\sqrt{3}}(1,1,-1)^\mathrm{T}$ for plane waves 1,2,3, and 4 respectively, as shown in Fig.~\ref{Fig3}a. This gives rise to a simple cubic lattice with lattice constant $a=\frac{\sqrt{
3}}{2}\lambda$. $\lambda$ is the wavelength of the plane wave. Experimentally, such superposition of multiple plane waves can be achieved with spatial light modulators.

\begin{table}
\centering
 \begin{tabular}{c c c} 
 \hline
 Plane wave  & $A_s$ & $A_p$ \\ 
 \hline
 1  & 0.0 & 1.0 \\ 
 2  & $1.0i\exp(-i\pi/15)$ & $-1.2\exp(-i\pi/15)$\\
 3  & $2.0i\exp(i\pi/10)$ & -$1.5\exp(i\pi/10)$\\
 4  & $2.0\exp(i\pi/5)$ & 0.0 \\
 \hline
 5 & 0.01 & 0.01 \\ [1ex] 
 \hline
 \end{tabular}
 \caption{Parameters for plane waves under superposition. Plane waves 1-4 are used to create the 3D spin defect lattice (Fig.~\ref{Fig3}). Plane waves 1-5 are used to create the quasi-periodic spin defect lattice (Fig.~\ref{Fig4}). The definition of polarization unit vectors are: $\mathbf{e}_s=\hat{\mathbf{k}}\times\hat{\mathbf{z}}/\sin\theta$, $\mathbf{e}_p=\mathbf{e}_s\times\hat{\mathbf{k}}$. $\theta$ is the angle between $\hat{\mathbf{k}}$ and $\hat{\mathbf{z}}$.}
 \label{sp_amp_table}
\end{table}

The spin density distribution inside a single unit cell is determined by the polarizations, amplitudes, and relative phases of the plane waves. Intuition may be drawn from two plane wave interference \cite{bekshaev2015transverse}. In our case, since the spin distributions on the opposite faces of the cubic unit cell are identical, the total topological charge inside one unit cell is zero. Therefore, a single unit cell contains an equal number of $+1$ and $-1$ topological charges. Here we demonstrate a configuration that gives rise to one pair of charges inside one unit cell. The amplitude choices are listed in Table \ref{sp_amp_table}. In Fig.~\ref{Fig3}b we show the topological charge distributions in $3\times 3\times 3$ unit cells. The positive and negative topological charges are denoted by red and blue dots. We see that each unit cell contains one pair of charges, and the defect points forms a 3D lattice in a form analogous to an ionic crystal.

In this example, at the topological spin defect point, the electric and magnetic field polarization is elliptical rather than linear. This is because the spin density have both the electric and the magnetic part, the individual contribution can be nonzero but the total spin density can be zero. Generically, the spin defect point does not coincide with polarization singularities, such as L-lines or V-points \cite{wang2021polarization}, since the polarization singularities are only determined by the electric fields. The example in Fig.~\ref{Fig1} of superposed Gaussian beams has its electric and magnetic field both being linearly polarized at the defect point is a special case.

In Fig.~\ref{Fig3}c we show the distribution of the spin inside one unit cell. The spin distributions are plotted on 5 planes with different values of $z$. Plane 2 and plane 4 intersects the spin defect point. Around such defect points we see the spin points to opposite directions on opposite side of the defect point. On plane 1, 3, and 5, we numerically calculate the topological flux $I_1$, $I_3$, and $I_5$ respectively, using $I = \int_\Sigma\mathcal{F}\mathrm{d}S$,
where $\mathcal{F}=\frac{1}{4\pi}\mathbf{n}\cdot\left(\frac{\partial\mathbf{n}}{\partial x}\times\frac{\partial\mathbf{n}}{\partial y}\right)$, $\Sigma$ is the two-dimensional (2D) unit cell in $xy$-plane. We obtain $I_1=I_5=0$, and $I_3=-1$. Given that plane 1 and 3 have a $+1$ topological charge between them, $I_1-I_3=1$. Similarly, $I_3-I_5=-1$. Moreover, not only the flux enclosing a defect point is quantized, the flux at each individual plane as indicated above is also quantized. This result come from the fact that $\pi_1(S^2)$ is trivial, thus the mapping from a 2D torus to $S^2$ has the same classification as the mapping from a 2D sphere to $S^2$ \cite{wojcik2020homotopy, han2017skyrmions}. Here the 2D torus refers to the 2D unit cell in the $xy$-plane, since the opposite edges in the unit cell has the same spin distribution and can be identified. Therefore, in the plane of $z=-0.06\lambda$ where plane 3 is located, we realized a 2D skyrmion lattice. Previously, a 2D skyrmion lattice of electromagnetic spin has been demonstrated for surface plasmon waves \cite{lei2021photonic, zhang2022optical}. Here we show that a 2D skyrmion lattice can also be obtained by interfering a few plane waves in free space. 

The existence of the topological spin defect points are quite general and does not involve fine tuning of the plane wave amplitudes and polarizations. However, the exact location of these defect points depend on parameters of the underlying plane waves. In Fig.~\ref{Fig3}d we show the location change of the topological spin defect points when we increase the amplitude of plane wave 3, simultaneously for both polarizations. The phases are unchanged. The trajectories of the defects are shown by the red and blue circles for the positive and negative charges, respectively. We see that the two charges move closer as we increase the amplitude. When the amplitude increase is $11.4\%$, the two charges meet each other at the location shown by the black dot in Fig.~\ref{Fig3}d and annihilate. No spin defect points is found upon further increase of such amplitude.

The spin defect points discussed here is analogous to what was previously discovered in magnetic materials, including individual defect points \cite{milde2013unwinding, da2014observation} and defect lattices \cite{kanazawa2016critical, fujishiro2019topological, okumura2020magnetic, kanazawa2020direct}. Achieving those spin textures in optical fields provides an alternative way to study such spin textures.

\section{Quasi-periodic spin defect lattices}
\begin{figure}
    \includegraphics[width=0.7\columnwidth]{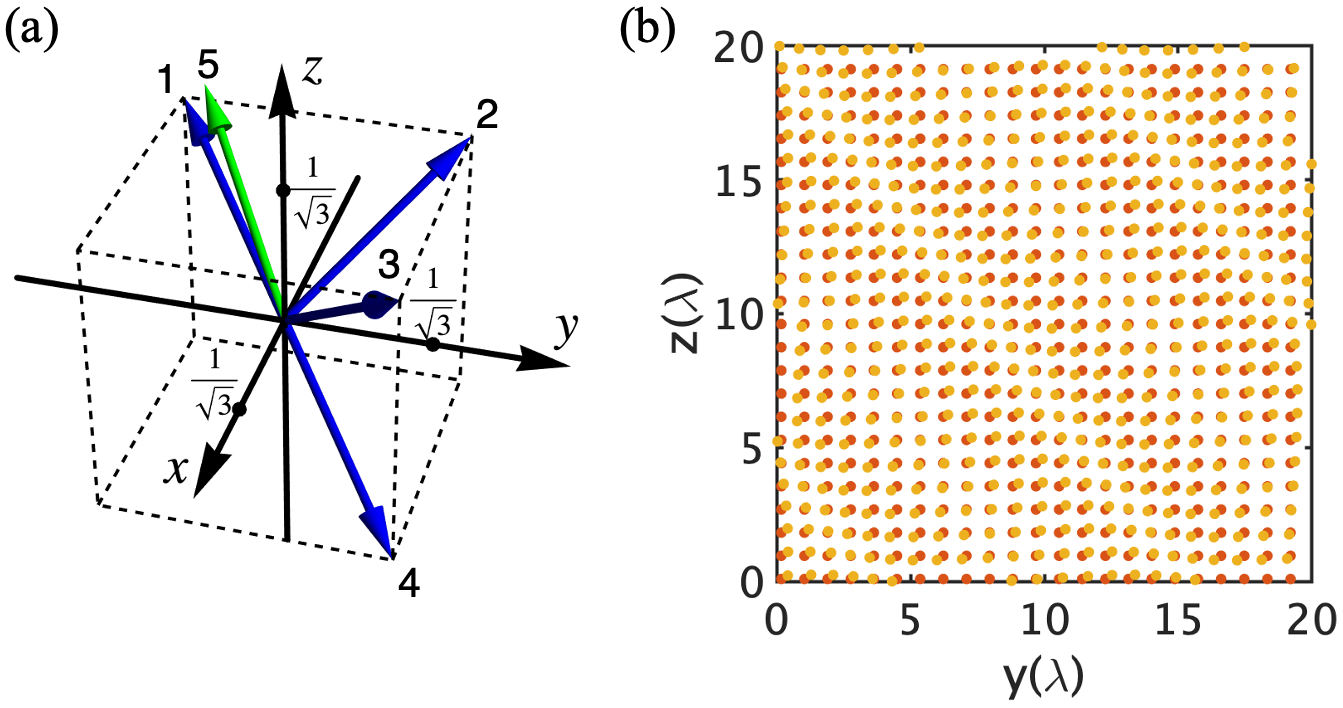}%
    \centering
    \caption{Quasi-periodic spin defect lattice in interfering plane wave systems. (a) The propagation direction $\hat{\mathbf{k_i}}$ for each plane wave under superposition. In spherical coordinates, plane wave 5 propagates at $\theta=5\pi/18$ relative to the $z$-axis, and $\phi=-7\pi/9$ relative to the $x$-axis. $\hat{\mathbf{k}}=\left(\sin\theta\cos\phi,\sin\theta\sin\phi,\cos\theta\right)$. (b) A 2D projection in the $yz$-plane of the quasi-periodic defect lattices, showing the quasi-periodic defect lattice (orange dots) and the 3D periodic defects in Fig.~\ref{Fig3} (red dots). Only the topological defects with positive charges in the range $-\sqrt{3}\lambda/4 < x < \sqrt{3}\lambda/4$ are shown. Displacements between the quasi-periodic lattice and the 3D periodic lattices are enlarged by ten times for better visualization. \label{Fig4}}
\end{figure}

Unlike magnetic materials, in our optical setup one can achieve more complex spin defect distributions. As an illustration, here we demonstrate that a quasi-periodic lattice of spin defect points can be formed by adding one more plane wave to the configuration discussed above. The directions of the wavevectors under superposition is shown in Fig.~\ref{Fig4}a. Plane wave 1-4 is the same as the example described above, whereas plane wave 5 (green arrow) is added to create a quasi-periodic lattice. Here we choose plane wave 5 to have a wavevector that is close to plane wave 1, with a relatively small amplitude (Fig.~\ref{Fig4}a and Table \ref{sp_amp_table}). Having such a plane wave 5 results in spatially dependent displacements from the original defect point locations created by plane waves 1-4. The displaced defects and the original defects are plotted in Fig.~\ref{Fig4}b, showing only the displacement in $y$ and $z$ directions. Strictly speaking, the displacements are not periodic. However, we can still identify periodicity in the displacement from the plot.

The periodic displacement can be understood as the interference length scale between plane wave 5 and its closest counterpart, plane wave 1. The periodicity can be calculated as:
\begin{equation}
    l_j = \frac{\lambda}{a_j|\hat{\mathbf{k}_{1j}}-\hat{\mathbf{k}_{5j}}|}
\end{equation}
Here $j=x,y,z$, $\hat{\mathbf{k}_{1j}}$ is the $j$-th component of $\hat{\mathbf{k}_{1}}$. $a_j$ is the periodicity along the $j$-direction that makes $l_j$ in the units of lattice sites. Calculation shows $l_x=121.9$, $l_y=13.6$, and $l_z=17.6$. The values of $l_y$ and $l_z$ match the periodicity shown in Fig.~\ref{Fig4}b.

\section{Discussion and conclusion}

Polarization is a concept closely related with the spin of light, but they are not the same. Polarization singularities carrying topological charges was discussed in literature \cite{bliokh2019geometric, wang2021polarization}. For 3D distribution of fields involving longitudinal components, the topological singularities are C-lines and L-lines, which correspond to lines of circular and linear polarizations, respectively. These lines carries topological charges given by 1D winding number, which is different from the point singularities of spin discussed here that carries 2D topological charges. Isolated polarization singularities (V-points) in 3D was also discussed \cite{kovalev2018tailoring, vyas2013polarization, otte2018polarization}, but along the propagation direction they generically splits into a pair of C-points and does not carry a 2D topological charge. Nontrivial topological configuration of polarization may exist in higher dimensional classifications. For example, the topological charge associated with the nontrivial $\pi_3(S^3)$ group is demonstrated \cite{sugic2021particle}.

Many applications can be envisioned with the topological spin defects of light. Since the spin state is different when displaced in different directions from the defect point, and the magnitude of spin also changes with increasing displacement, one may infer the position of a subwavelength particle by analyzing the light it scatters \cite{beckley2010full, neugebauer2018magnetic}. This may allow one to achieve 3D tracking of subwavelength particles with subwavelength resolution.

The spin of light also has mechanical effects. The torque of a sufficiently small particle inside monochromatic light field is given by \cite{canaguier2013force, bliokh2014magnetoelectric, bekshaev2015transverse}:
\begin{equation}
    \begin{split}
    \mathbf{T}&=\mathrm{Re}(\alpha_e\mathbf{E^*}\times\mathbf{E}+\alpha_m\mathbf{H^*}\times\mathbf{H}) \\
    &\propto \mathrm{Im}(\alpha_e)(\mathbf{S}_e+\frac{\mathrm{Im}(\alpha_m)}{\mathrm{Im}(\alpha_e)}\mathbf{S}_m)\\
    \end{split}
\end{equation}
Here $\alpha_e$ and $\alpha_m$ are electric and magnetic polarizabilities of the particle. $\mathbf{S}_e=\frac{1}{4\omega}\epsilon_0\mathrm{Im}(\mathbf{E^*}\times\mathbf{E})$ and $\mathbf{S}_m=\frac{1}{4\omega}\mu_0\mathrm{Im}(\mathbf{H^*}\times\mathbf{H})$ are electric and magnetic part of the spin density, respectively. Instead of the total spin $\mathbf{S}$, the quantity $\mathbf{S}_e+\frac{\mathrm{Im}(\alpha_m)}{\mathrm{Im}(\alpha_e)}\mathbf{S}_m$ determines the distribution of torque over space, and may host topological singularities similar to the one we have described above. We note that here both $\mathrm{Im}(\alpha_m)\neq0$ and $\mathrm{Im}(\alpha_e)\neq0$ are crucial to achieve isolated singularities of torque. The latter condition is generically true for almost all particles, therefore we only discuss $\mathrm{Im}(\alpha_m)$ hereafter. If $\mathrm{Im}(\alpha_m)=0$, the torque will be proportional to $\mathbf{S}_e$ only. Points with $\mathbf{S}_e=0$ has linearly polarized electric fields and are known to form lines in 3D, known as L-lines \cite{berry2001polarization, freund2010optical} (not to be confused with L-surfaces when only paraxial components are considered \cite{flossmann2005polarization}). In this case, points where the torque is zero form lines rather than points. We show the torque configuration under different $\mathrm{Im}(\alpha_m)/\mathrm{Im}(\alpha_e)$ values in the supplementary section 4. When $\mathrm{Im}(\alpha_m)\neq0$, we obtain isolated singularities of torque along with topological configuration of torque around such singularities. Such topological torque singularities allow one to conveniently apply arbitrary rotation axis to the particle.

In conclusion, we have shown electromagnetic field in 3D space hosts topological spin defects. These defects carries topological charge manifested as nontrivial spin texture on the sphere surrounding such defect point. Due to their topological nature, the existence of such defects are robust under perturbations. We demonstrated examples that create isolated spin defect point, and periodic or quasi-periodic array of spin defect points. Given this is a topological phenomena, spin defect points will not be limited to the wave forms discussed here and can be found in electromagnetic waves in general. Compared to their analog in magnetic materials, shaping the distribution or location of those defect points will be relatively easy in optics. We envision such topological spin defects will find applications in 3D imaging and nanoparticle manipulation.

\section{Backmatter}


\begin{backmatter}
\bmsection{Funding}
The work is supported by the U. S. Office of Naval Research (Grant No. N00014-20-1-2450), and by a Simons Investigator in Physics grant from the Simons Foundation (Grant No. 827065).

\bmsection{Disclosures}
The authors declare no conflicts of interest.

\bmsection{Data Availability Statement}
No data were generated or analyzed in the presented research.

\bmsection{Supplemental document}
See Supplement 1 for supporting content. 

\end{backmatter}

\bibliography{ref}

\end{document}